\documentclass[final,5p,times,twocolumn]{elsarticle}
\usepackage{amsmath,amsfonts,amssymb,upgreek}
\usepackage{algorithmic}
\usepackage{algorithm}
\usepackage{array}
\usepackage[caption=false,font=normalsize,labelfont=sf,textfont=sf]{subfig}
\usepackage{textcomp}
\usepackage{stfloats}
\usepackage{url}
\usepackage{verbatim}
\usepackage{graphicx,svg}
\usepackage{cite}
\usepackage[hidelinks]{hyperref}
\usepackage{float}

\newcommand{\neff}{n_{\text{eff}}}

\begin{document}

\begin{frontmatter}

\title{Linearly Multiplexed Photon Number Resolving Single-photon Detectors Array}

\author[UniTN,FBK]{Leonardo Limongi}

\affiliation[UniTN]{
    organization={Department of Industrial Engineering, University of Trento},
    addressline={Via Sommarive 9},
    city={Povo},
    postcode={38123},
    state={Trento},
    country={Italy}
}
\affiliation[FBK]{
    organization={Center for Sensors and Devices, Bruno Kessler Foundation},
    addressline={Via Sommarive 18},
    city={Povo},
    postcode={38123},
    state={Trento},
    country={Italy}
}

\author[IFN]{Francesco Martini}

\affiliation[IFN]{
    organization={Istituto di Fotonica e Nanotecnologie, CNR-IFN},
    addressline={Via del Fosso del Cavaliere 100},
    postcode={00133},
    state={Roma},
    country={Italy}
}

\author[UniRoma2,INFNRoma2]{Thu Ha Dao}

\affiliation[UniRoma2]{
    organization={Department of Industrial Engineering, University of Rome Tor Vergata},
    addressline={Via del Politecnico 1},
    postcode={00133},
    state={Roma},
    country={Italy}
}
\affiliation[INFNRoma2]{
    organization={Istituto Nazionale di Fisica Nucleare, Sezione di Roma Tor Vergata},
    addressline={Via della Ricerca Scientifica},
    postcode={00133},
    state={Roma},
    country={Italy}
}

\author[IFN]{Alessandro Gaggero}

\author[UniTN]{Hamza Hasnaoui}

\author[UniTN,FBK]{Igor Lopez-Gonzalez}

\author[IFN]{Fabio Chiarello}

\author[UniRoma2,INFNRoma2]{Fabio De Matteis}

\author[UniTN,TIFPA]{Alberto Quaranta}

\affiliation[TIFPA]{
    organization={INFN TIFPA, Trento Institute for Fundamental Physics and Applications},
    addressline={Via Sommarive 14},
    city={Povo},
    postcode={38123},
    state={Trento},
    country={Italy}
}

\author[INFNRoma2]{Andrea Salamon}

\author[IFN]{Francesco Mattioli}

\author[FBK]{Martino Bernard}

\author[UniTN,TIFPA]{Mirko Lobino}

\begin{abstract}
Photon Number Resolving Detectors (PNRDs) are devices capable of measuring the number of photons present in an incident optical beam, enabling light sources to be measured and characterized at the quantum level. In this paper, we explore the performance and design considerations of a linearly multiplexed photon number-resolving single-photon detector array, integrated on a single mode waveguide. Our investigation focus on defining and analyzing the fidelity of such an array under various conditions and proposing practical designs for its implementation. Through theoretical analysis and numerical simulations, we show how propagation losses and dark counts may have a strong impact on the performance of the system and highlight the importance of mitigating these effects in practical implementations.
\end{abstract}

\begin{keyword}
Photon Number Resolving Detector, Single-photon Detector, Superconducting Nanowire Single-photon Detector, Lithium Niobate-on-Insulator, Niobium Nitride, Photonic Integrated Circuit, Photonic chip
\end{keyword}

\end{frontmatter}

\section{Introduction}
\label{sec:Introduction}

Photon Number Resolving Detectors (PNRDs) are devices capable of measuring the number of quanta of energy present in an incident optical field, enabling light sources to be measured and characterized at the quantum level \citep{2019Helversen, 2023Stasi, 2005OHadfield}. PNRDs have a pivotal role in various optical applications and quantum technologies. In quantum key distribution \citep{2020Xu}, PNRDs can be used in specific protocols \citep{2018Cattaneo} to guarantee secure transmission of information. In quantum imaging, they enable high-resolution by precisely counting the photons emitted or scattered by objects of interest \citep{2022Wolley}. Moreover, they can be used in other applications such as quantum state preparation \citep{2016Harder, 2022Kuts}, photon-counting Lidar \citep{2014Huang}, quantum metrology \citep{2019Wu}, and quantum tomography \citep{2019Olivares}.

Photon Number Resolving Detectors can be realized with different technologies, each one with its characteristics, operating principles, and limitations. For instance, schemes with photomultiplier tubes \citep{2004Zambra} and Geiger-mode avalanche photodiodes (GAPDs) \citep{2004Blazej} have shown photon number resolution with few photons. Visible light photon counters \citep{2003Waks} have a high dark count rate and slow operational rate, while transition edge sensors (TES) can discriminate the number of photons with high efficiency (98\%) \citep{2011Fukuda} and low noise at telecommunication wavelengths, but their performance is limited by the low count rates and high time jitter \citep{2003Miller, 2008Lita}.

Superconducting Nanowire single photon detector (SNSPD) \citep{2012Natarajan} is considered one of the most promising platform for overcoming the limitations of the detector platforms mentioned above, thanks to its high efficiency \citep{2020Reddy}, negligible dark count rate \citep{2015Shibata}, jitter time in the picosecond region, \citep{2020Korzh} and short recovery time \citep{2021Perrenoud}. However, single SNSPD detectors do not have intrinsic photon number resolution capability like TES detectors. Here we show how this limit can be addressed with a multiplexed scheme on an optical waveguide to enable photon number resolution and optimized fidelity.

Integration of SNSPD detectors on optical waveguide chips has been demonstrated, showing how these devices maintain their outstanding properties for a diverse range of materials \citep{2021Esmaeil}. Integration with complex photonic waveguide networks has shown the possibility of creating a combined platform which contains both capabilities of SNSPDS and PNRDs with great scalability potential \citep{2021Martini, 2023Cheng, 2019Gaggero, 2009Hu, 2015Najafi} and applications in quantum communication \citep{2023Notarnicola}, characterization of quantum photon statistics \citep{2023Cheng} and quantum sensing \citep{2021Endo}.

Here we propose a design for a linearly multiplexed PNRD based on SNSPD, integrated on a single optical waveguide. 
First, keeping generality, we show the theoretical calculations of each element efficiency to maximize the fidelity, accounting for waveguide losses and dark counts. Then, a design is proposed, showing a guideline to obtain optimized parameters considering fabrication constraints.
For our design we focus on silicon nitride strip loaded waveguide on a thin film lithium niobate (TFLN) substrate \citep{2022Peace} as photonic platform because it offers the great potential of TFLN in terms of nonlinearty and electrooptical reconfigurability, but it is more agile to fabricate as it exploits the well established manufacturing techniques for silicon nitride leaving the hard-to-process LN film intact.

This proposal aims at the integration of PNR capability in a photonic integrated circuit (PIC) that, thanks to TFLN properties, can exploit the nonlinear properties of lithium niobate to generate and manipulate quantum states of light \citep{2018Lenzini} on a single chip. 
Moreover, due to its linearity, our design can be considered modular in the sense that SPD elements can be added to the detector array incrementally. 

The manuscript is structured as follows:
In Sect.~\ref{sec:Performance} we will discuss a general approach for the geometry of an array of SNSPDs on a linear waveguide and calculate the relevant merit factors in the case of ideal waveguide and detectors.  
Then, non-idealities such as propagation losses and detector dark counts will be discussed and estimated. 
In Sect.~\ref{sec:Design} an actual design for the implementation of the waveguide PNRD will be proposed and discussed. Simulations will be used to calculate the relevant parameters for the proposed device so that the final merit factors can be calculated using the theory developed in Sect.~\ref{sec:Performance}. In Section IV we will summarize the results and draw our conclusions. 

\section{A multiplexed SNSPDs on a single waveguide}
\label{sec:Performance}

\subsection{Lossless waveguide and detector}
\begin{figure}
    \centering
    \includegraphics{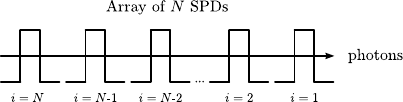}
    \caption{Photon-number resolving detector array geometry: photons travel from left to right through a series of $N$ single-photon detectors, indexed from $N$ to 1.}
    \label{fig:PNRD_geometry}
\end{figure}

First, we consider the ideal case of an array of $N$ single-photon detectors (SPDs) coupled to a lossless waveguide, with no dark counts, and schematically shown in Fig.\ref{fig:PNRD_geometry}. Each element of the array works in the Geiger mode and can only discriminate between light and no light, generating the same signal for any number of absorbed photons. The detectors are numbered so that the first element to be illuminated has index $N$ and the last is element $1$.
The fidelity $F$ in this ideal system as a function of the number of incoming photons $m$ and the number of SPDs $N$ is given by the probability of correctly detecting all of the $m$ incoming photons, expressed as
\begin{equation}
F=p(\text{detected}=m|\text{incoming}=m).
\label{Eq:FidelityProbabilisticDefinition}
\end{equation}
The probability in Eq.~\ref{Eq:FidelityProbabilisticDefinition} can be reinterpreted as the probability that each element of our array absorbs and detects either zero or one photon and no photon is lost at the end of the array. 

To derive its expression, consider $N$ SPDs arranged linearly as depicted in Fig.~\ref{fig:PNRD_geometry} and let $\mathcal{D}=\left\{i\right\}_{i=1}^{i=N}$ denote the ordered set of detector indices with the value $N$ associated to the first detector illuminated and $1$ to the last. Define $\mathcal{F}\subset\wp\left(\mathcal{D}\right)$ as the family of sets containing $m$ elements in the power set of $\mathcal{D}$, representing all possible combinations of $m$ different detectors detecting the $m$ incoming photons. Recall that the power set of $\mathcal{D}$ is the set of all subsets of $\mathcal{D}$, namely
\begin{equation*}
\begin{aligned}
\wp\left(\mathcal{D}\right) =& \left\{\mathcal{S}\;\big|\;\mathcal{S}\subseteq\mathcal{D}\right\}=\\
=&\big\{\varnothing,
\{1\},\{2\},\dots,\{N\},\{1,2\},\{1,3\},\dots,\\
&\hspace{1.4cm}\{N-1,N\},\dots,\{1,2,\dots,N\}
\big\}
\end{aligned}
\end{equation*}
and that $|\wp\left(\mathcal{D}\right)|=2^N$ is the number of elements inside $\wp\left(\mathcal{D}\right)$.\\
Hence, $\mathcal{F}$ can be expressed as
\begin{equation*}
\begin{aligned}
\mathcal{F} =& \left\{\mathcal{S}\in\wp\left(\mathcal{D}\right)\;\big|\;|\mathcal{S}|=m\right\}=\\
=& \big\{\{1,2,\dots,m-1,m\},\{1,2,\dots,m-1,m+1\},\dots,\\
&\hspace{3.25cm}\{N-m,N-m+1,\dots,N\}\big\}
\end{aligned}
\end{equation*}
which contains $\binom{N}{m}=\tfrac{N!}{m!(N-m)!}$ elements, or  $m$-dimensional vectors denoted as $\vec{k}$ equivalently, identifying every possible choice of $m$ indices out of the total $N$ of detectors. It is important to notice that $\mathcal{F}$ contains every possible list of SPDs that have detected one photon with the constraint of both no photon losses and at most one photon per SPD.

Now, denote $\eta_i$ as the efficiency of the $i$-th detector to click when illuminated with a single photon, we can write the probability of a photon being absorbed by the $i$-th element of the array as
\begin{equation}
p_i = \eta_i\prod_{j=i+1}^{N+1}\left(1-\eta_j\right)
\label{Equation:ProbabilityFactors}
\end{equation}
where $\eta_{N+1}=0$, representing free propagation before entering the array, and the product $\prod_{j=i+1}^{N+1}\left(1-\eta_j\right)$ indicates the probability of a photon not being absorbed before the $i$-th detector.

Under these assumptions, the probability of distributing at most one of the $m$ incoming photons over the $N$ SPDs, or equivalently the fidelity of the PNR detector, is given by
\begin{equation}
F(m,N)=\sum_{\vec{k}\in\mathcal{F}}m!\prod_{k_i\in\vec{k}}\eta_{k_i}\prod_{j=k_i+1}^{N+1}(1-\eta_j)
\label{Equation:Fidelity_with_etas}
\end{equation}
where $m!$ accounts for all permutations of the photons inside each $\vec{k}$ vector.

When there are $m$ incoming photons, the maximum fidelity is obtained when the probability of absorbing a single photon is uniform across each element of the array \citep{2019Jonsson} and is given by 
\begin{equation}
    p_i=\frac1N\ \forall i\in[1,N].
    \label{eq:uniform_prob}
\end{equation}
Substituting Eq.~\ref{eq:uniform_prob} in Eq.~\ref{Equation:ProbabilityFactors}, the efficiency of the $i$-th element of the array is given by
\begin{equation}
\eta_i=\frac{1}{i}\ \ \text{with}\ \ 1\leq i\leq N
\label{Equation:eta_i}
\end{equation}
Equation~\ref{Equation:eta_i} implies that the first detector of the array will have the smallest efficiency of $1/N$ while the last one requires total absorption ($\eta$=1) for preventing light from escaping the array, corresponding to a nanowire of infinite length.
Combining this result with Eq.~\ref{Equation:Fidelity_with_etas}, we obtain a maximum fidelity for the detection of $m$ photons with an array of $N$ detectors of
\begin{equation}
F(m,N)=m!\binom{N}{m}\frac{1}{N^m}.
\label{Equation:Ideal_Fidelity}
\end{equation}

\begin{figure}[t]
    \centering
    \includegraphics[scale=0.5]{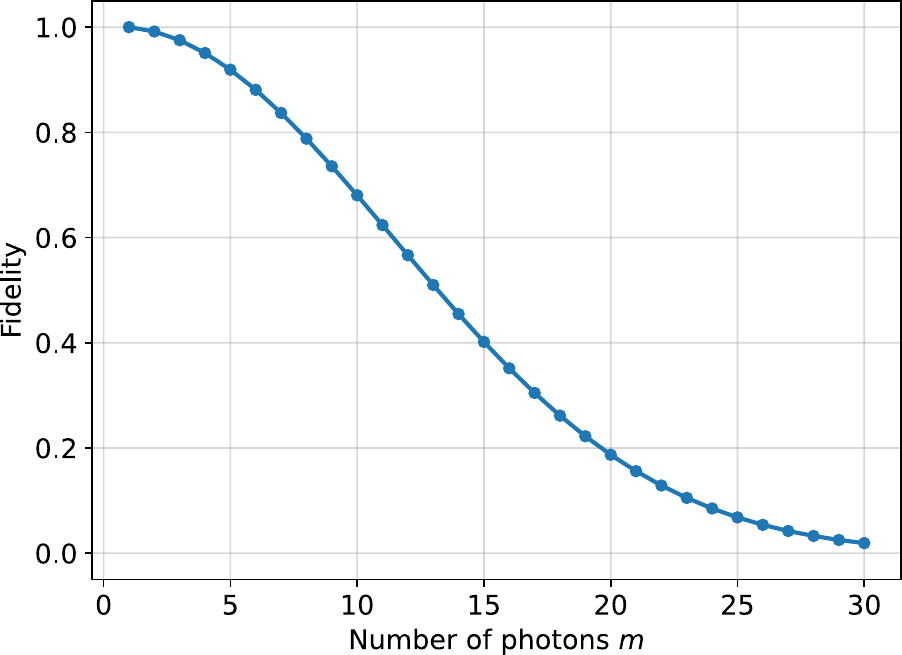}
    \caption{Fidelity versus the number of incoming photons $m$ (blue points) with $N=120$ single-photon detectors. This plot was obtained by a second order Stirling expansion of Eq.~\ref{Equation:Ideal_Fidelity}.}
    \label{Figure:Fidelity_on_photons}
\end{figure}

It is worth to note that the same result is obtained in \citep{2020Zou} by starting from the same assumptions.
The fidelity of Eq.~\ref{Equation:Ideal_Fidelity} is shown in Fig.~\ref{Figure:Fidelity_on_photons} for an array of $N=120$ elements, and it represents the ideal case in which we considered only dichotomous events of photons being either absorbed or not.

\subsection{Propagation losses and fidelity}
All real waveguide suffers from propagation losses, and in this situation Eq.~\ref{Equation:Ideal_Fidelity} is not valid. In this case the ratio between the losses introduced by the waveguides and the absorption of the SPDs plays a fundamental role for the calculation of the fidelity.

Consider a waveguide with a loss coefficient $\alpha_w$, and a superconducting nanowire with absorption coefficient $\alpha_c$. As the mechanisms are the same for both the elements, the probability of a photon being absorbed or not over a distance $L$ is given by:
\begin{equation}
\mathcal{A} = 1-e^{-\alpha L}, \qquad \mathcal{T} = 1-\mathcal{A} = e^{-\alpha L}
\label{Equation:AbsorptionTransmissionRelations}
\end{equation}
where $\mathcal{A}$ and $\mathcal{T}$ are, respectively, the absorption and the transmission probabilities, both governed by the coefficient $\alpha$. For the i-th element of the array Eq.~\ref{Equation:eta_i} holds and we have $\mathcal{A}_i=\eta_i=1/i$, from where we can extract the length of the element by inverting the first of Eqs.~\ref{Equation:AbsorptionTransmissionRelations} with absorption coefficient $\alpha_c$, namely
\begin{equation}
L_i = \frac{1}{\alpha_c}\ln\left(\frac{1}{1-\mathcal{A}_i}\right) = \frac{1}{\alpha_c}\ln\left(\frac{i}{i-1}\right).
\label{Equation:Length_efficiency_relation}
\end{equation}

To incorporate propagation losses into the fidelity computation, we modify Eq.~\ref{Equation:ProbabilityFactors} for the probability of a photon being absorbed by the i-th detector as follows
\begin{equation}
p_i = \mathcal{A}_i^{(\text{nw})}\prod_{j=i+1}^{N+1}\mathcal{T}_j^{(\text{nw})}\mathcal{T}_j^{(\text{wg})}
\label{Equation:ProbabilityFactorsWithLosses}
\end{equation}
where we assume photons to survive both the waveguides (wg) and the nanowires (nw) until before the $i$-th element and being absorbed by it.

Substituing Eq.~\ref{Equation:ProbabilityFactorsWithLosses} into Eq.~\ref{Equation:Fidelity_with_etas}, and leveraging Eqs.~\ref{Equation:AbsorptionTransmissionRelations} and \ref{Equation:Length_efficiency_relation}, we derive the following loss-perturbed fidelity expression
\begin{equation}
F(m,N,r)=m!\left(\frac1N\right)^{m\left(r+1\right)}\sum_{\vec{k}\in\mathcal{F}}\prod_{k_i\in\vec{k}}k_i^r
\label{Equation:Lossy_Fidelity}
\end{equation}
where $r=\alpha_w/\alpha_c$ represents the ratio of the absorption coefficients of the waveguide and the nanowire detector.

\begin{figure}[t]
    \centering
    \includegraphics[scale=0.5]{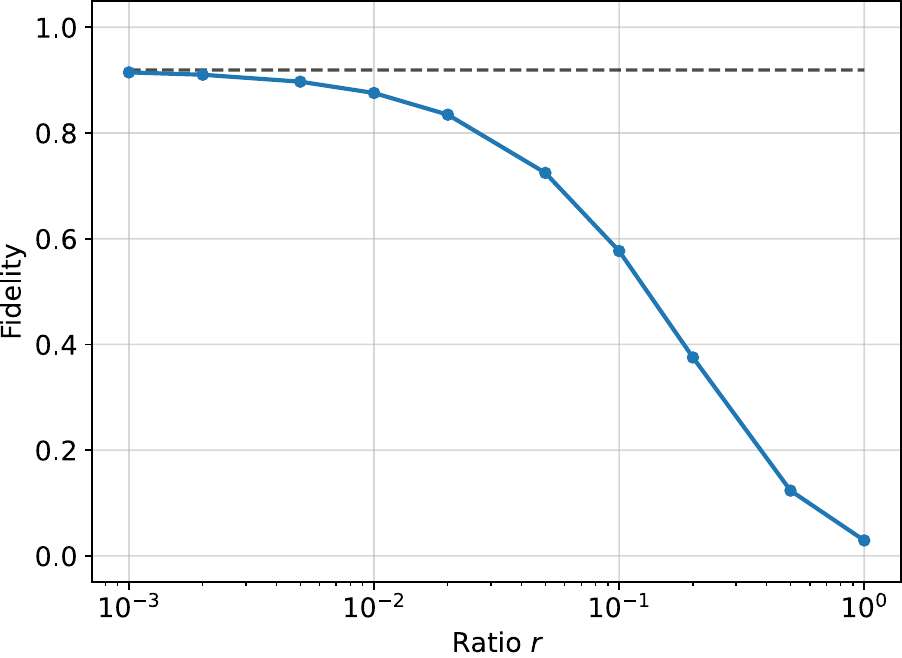}
    \caption{Fidelity with progation losses versus the ratio $r=\alpha_w/\alpha_c$ (blue points - Lossy) with $N=120$ single-photon detectors and $m=5$ incoming photons. For an increasing $r$ the fidelity drops from the lossless of 0.9191 (black dashed line) to zero in a monotonic and continuous way.}
    \label{Figure:Fidelity_on_ratio}
\end{figure}

Figure~\ref{Figure:Fidelity_on_ratio} shows how fidelity changes with respect to $r$ when we send 5 photons into an array of 120 detectors, each with length given by Eq~\ref{Equation:Length_efficiency_relation}.

\subsection{Detector with dark counts}
Next we introduce detectors dark counts in our calculation assuming a dark count rate (DCR) per unit time  proportional to the length of the nanowire \citep{2015Engel}. For an array of $N$ detectors where the length of each element is given by Eq.~\ref{Equation:Length_efficiency_relation}, the total length $L_{\text{tot}}$ is
\begin{equation}
    L_{\text{tot}} = L_1 + \sum_{i=2}^{N}L_i = L_1 + \frac{1}{\alpha_c}\ln N,
\end{equation}
where we isolated the divergent term for $i=1$ because in the ideal case the last detector absorbs 100\% of the photons and have infinite length.

We assume a $\text{DCR}=L_{\text{tot}}\gamma$, where $\gamma$ is a coefficient that depends on the properties of the superconducting nanowire. The probability of having $k$ dark counts in a time interval $\Delta t$ follows the Poisson distribution
\begin{equation}
p\left(k,\Delta t\right)=\frac{\left(L_{\text{tot}}\gamma \Delta t\right)^k}{k!}e^{-\left(L_{\text{tot}}\gamma \Delta t\right)}
\label{Eq:DCRprobability}
\end{equation}

The probability of having no dark counts for the measurement time $\Delta t$ is given by $p(0,\Delta t)$, and hence we can multiply it to Eq.~\ref{Equation:Lossy_Fidelity} to obtain the loss-DCR-perturbed fidelity
\begin{equation}
\begin{aligned}
F(m,N,&r,\gamma \Delta t)=\\
&m!\left(\frac1N\right)^{m\left(r+1\right)}N^{-\frac{\gamma \Delta t}{\alpha_c}}e^{-\gamma \Delta t L_1}\sum_{\vec{k}\in\mathcal{F}}\prod_{k_i\in\vec{k}}k_i^r
\end{aligned}
\end{equation}
For state of the art SNSPDs this correction is usually quite small since the dark count rate can go down to few tens of Hz and the measurement time is of the order of ns \citep{2019Gaggero}.

\section{Design of the detector array on a strip-loaded waveguide on TFLN}
\label{sec:Design}
\begin{figure}[t]
    \centering
    \includegraphics{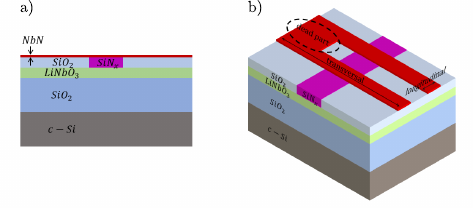}
        \caption{Cross sections of the photon-number resolving detector: starting from the bottom there are a crystalline silicon ($c$-Si) layer, a silicon oxide (SiO$_2$) layer, a lithium niobate (LiNbO$_3$) layer, a silicon nitride (SiN$_x$) waveguide surrounded by an SiO$_2$ layer, and finally the niobium nitride (NbN) layer. The detailed dimensions are reported in Tab.\ref{tab:Parameters}.}
    \label{fig:Cross_Section}
\end{figure}

The relations for the fidelity derived in the previous sections can be applied to any material platform for either the waveguide and the nanowire detectors. Now we consider a device fabricated on a lithium niobate on insulator (LNOI) substrate loaded with a Silicon Nitride (Si$_3$N$_4$ or SiN) waveguide strip and coupled to a series of Niobium Nitride (NbN) SNSPDs, with the geometry sketched in Fig.~\ref{fig:Cross_Section}a.
The PNRD is fabricated on a LNOI substrate made by a $300$~nm lithium niobate layer on a 3~$\mu$m silicon dioxide (SiO$_2$) insulator. On the LNOI film, a ridge in SiN is patterned with a nominal a width of 1.2~$\mu$m and a thickness of 200~nm, defining the waveguide channel. The SNSPDs are then patterned on top of the SiN ridge with a fixed thickness of $6$~nm and a width of 80~nm.
Such series of SNSPDs can be realized on top of the strip-loaded LNOI photonic circuit thanks to the following fabrication steps. A SiO$_2$ cladding layer is deposited by Plasma-Enhanced Chemical Vapor Deposition, and a planarization step is performed by Chemical Mechanical Polishing (CMP). The NbN superconducting film is deposited using DC Magnetron Sputtering, and the different nanowire patterns are realized by Electron Beam Lithography followed by dry etching. AuPd resistors, placed in parallel to each nanowire element, and Au contact pads are fabricated via lift-off. Such fabrication procedure is the same employed in \citep{2013Sahin}, except for the CMP step that allows to position both the contact pads and AuPd parallel resistors on the cladding layer. In addition, such step grants optimal surface roughness prior the SNSPD fabrication.\\
Such a structure is simulated using the finite elements method (FEM) with the parameters summarized in Tab.~\ref{tab:Parameters}.
\renewcommand{\arraystretch}{2.4} 
\begin{table}[h]
    \centering
    \begin{tabular}{||c|c|c|c|c||}
        \hline
        $\lambda$ = 1.55~$\mathrm{\mu m}$\ & SiO$_2$ & LiNbO$_3$ & Si$_3$N$_4$ & NbN \\
        \hline
        Refractive index $n$ & 1.44 &
        $\begin{gathered}
            2.21 (o)\\
            2.14 (e)
        \end{gathered}$
        & 1.88 & $\begin{gathered}
            5.23 +\\
            i5.82
        \end{gathered}$ \\
         Thickness $h$ $\left(\mathrm{\mu m}\right)$ & $3.00$ & $0.3$ & $0.2$ & $0.006$ \\
         Width $\left(\mathrm{\mu m}\right)$ & slab & slab & $1.20$ & $0.08$ \\
        \hline
    \end{tabular}
    \vspace{5pt}
    \caption{List of parameters used for the  FEM simulations. Both SiO$_2$ and SiN refractive index are obtained by Variable Angle Spectroscopic Ellipsometry measurements on fabricated samples. LiNbO$_3$ and NbN refractive index are respectively reported in \citep{1997Zelmon} and \citep{2008Anant}.}
    \label{tab:Parameters}
\end{table}

\subsection{Straight and curved propagation}
Initially, to estimate the propagation losses in passive structures, and determine $\alpha_w$, FEM simulations were conducted for strip-loaded waveguides, excluding the NbN layer, for both straight and curved configurations.

For the first part of simulations, we performed a parametric sweep varying both the SiN waveguide width and height to take into account different fabrication scenarios. The waveguide height ranged from 180 to 220~nm, while the width varied between 1~$\mathrm{\mu m}$ and 1.4~$\mathrm{\mu m}$.
For a thickness of 200~nm and a width of 1.2~$\mu$m, we obtained a single mode waveguide with an effective refractive index $n_{\text{eff}}=1.8066$.

After determining the single mode $n_{\text{eff}}$ under ideal conditions, we looked at the stability of such waveguides with respect to slab leakages. The latter can be characterized by the difference in the effective refractive index between strip-loaded and slab modes \citep{2019Boes}, denoted as $\Delta n = n_{\text{eff}}^{(\text{strip})}-n_{\text{eff}}^{(\text{slab})}$. Following the last reference, a simple way to see whether or not lateral leakage occurs, is to study the sign of $\Delta n$. Indeed, strip-loaded waveguides suffer from strong lateral leakage when $\Delta n < 0$. 

For this purpose, we modified the simulation setup by removing the SiN waveguide layer, leaving only the LN slab with an additional oxide layer on top. To ensure comparability with previous results, we conducted a parametric sweep on the thickness of the oxide layer using the same values as in the strip-loaded waveguide simulations. In our scenario, both TE and TM polarizations of the slab modes exhibit effective refractive indices ($\neff^{(\text{slab,TE})}=1.7714$ and $\neff^{(\text{slab,TM})}=1.6070$ respectively) lower than that of the strip-loaded TE mode, thus providing robust mode isolation during straight propagation with respect to lateral leakages.

\begin{figure}
    \centering
    \includegraphics[scale=.5]{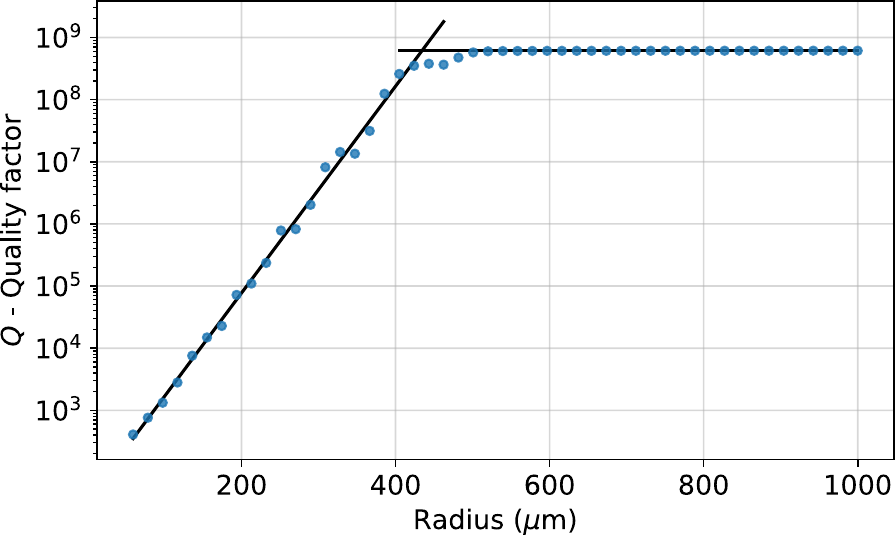}
    \caption{Log-scale plot of the quality factor of the simulated ring resonator as a function of the radius. The blue points represent the simulation results, while the black lines are the fit for both horizontal behaviors, that saturates at $Q_\infty=6.11\cdot10^8$, and the slope. The latter has a value of $1.66\cdot10^{-1}$~dB/$\mu$m and it indicates how much the quality factor's order of magnitude changes with respect to the bending radius. The intersection of the fits is at $R=434~\mu\text{m}$, representing the radius under which the bending radiative losses are the dominating factor.}
    \label{fig:Quality_vs_radius}
\end{figure}

In curved waveguides, radiative leakage rises sharply when the bending radius $R$ is sufficiently small. To assess this possibility and determine the critical bending radius $R_{\text{crit}}$ at which radiative losses become significant, we conducted FEM simulations of a strip-loaded ring resonator. Maintaining constant waveguide thickness and width, we evaluated the quality factor of the ring as a function of its radius, ranging from 40~$\mathrm{\mu m}$ to 1~mm. These results are shown in Fig.~\ref{fig:Quality_vs_radius} which exhibits two distinct regions. First, for small radii, there is an exponential decay of the ring quality factor with decreasing radii which indicates that radiative losses dominate over other loss mechanisms. In the the second region, for larger radii, we see a saturation of the ring quality factor to a value of $6.11\cdot10^8$ which suggests that bend-related radiative losses become negligible beyond a certain radius. The remaining sources of losses could be either physical, such as substrate losses or scattering due to the waveguide roughness, or non-physical, for example due to artifacts of the simulation such as the finite size of the simulation window.

From the intersection of the linear fits (on the y-log scale) of the data in Fig.~\ref{fig:Quality_vs_radius}, we estimate a critical radius $R_{\text{crit}}$=434~$\mathrm{\mu m}$. Below $R_{\text{crit}}$ where radiative losses dominate, the slope of the fit is 0.166~$\frac{\text{dB}}{\mathrm{\mu m}}$ and indicates the quality factor's order of magnitude variation with respect to the variation of the bending radius.

\subsection{Tuning SPDs efficiency}
\begin{figure}[h]
    \centering
    \includegraphics{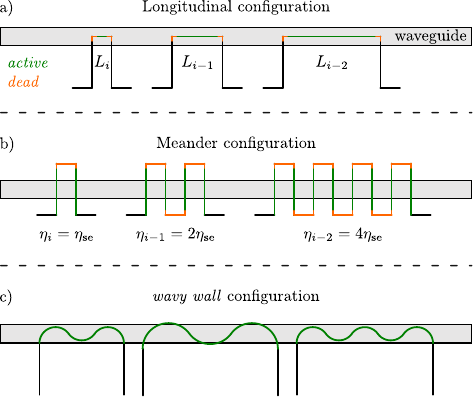}
    \caption{Different disposition of the NbN nanowire along the waveguide with highlighted \textit{active} and \textit{dead} parts respectively in orange and green colors. Panel a) shows the longitudinal configuration, characterized by having the detecting NbN nanowire parallel to the waveguide, and the detection efficiency can be continuously tuned by varying the nanowire's length $L_i$. Panel b) shows the meander configuration, characterizing by the detecting nanowire being perpendicular to the waveguide. Here the efficiency is discretized, and can be selected only among multiples of the single-element efficiency $\eta_{\text{se}}$. Panel c) shows the \textit{wavy wall} configuration, being characterized by having the active part of the nanowire in a periodic and uniform-curvature shape. Specifically, each SPD is formed by circumference arcs with fixed radius.}
    \label{fig:Different_configurations}
\end{figure}

The geometry of NbN nanowires along the waveguide is a key aspect for designing a good PNRD.
Figure~\ref{fig:Different_configurations} reports a summary of the geometry we investigated. Advantages and drawbacks of each configuration are discussed in this section.

As stated in the previous sections, the absorption efficiency of a nanowire can be tuned by its interaction length with the mode of the waveguide. For such a purpose, disposing the nanowire parallel to the waveguide, as showed in Fig.~\ref{fig:Different_configurations}a, could be the best solution for the fabrication of efficiency-tunable NbN SPDs. Indeed, our simulations show that a nanowire in such a configuration absorbs $\alpha_c = 422$~$\frac{\text{dB}}{\text{cm}}$, while SiN strip-loaded LNOI waveguides can reach propagation losses as low as $\alpha_w=0.3$~$\frac{\text{dB}}{\text{cm}}$ \citep{2018Boes}. This number translates into a ratio $r<10^{-3}$ and a fidelity close to the ideal case as shown in Fig.~\ref{Figure:Fidelity_on_ratio}.\\
Additionally, one can estimate the probability of having zero dark counts per measurement (Eq.~\ref{Eq:DCRprobability}) as follows: suppose $N=120$ SPDs in the array, the worst scenario reported in \citep{2019Gaggero} suggests considering $\text{DCR}=30k\text{~Hz}$ for a 60~$\mu$m nanowire, the measurement time can be taken as $t=10$~ns as it shall not be less than the latching time of SNSPDs \citep{2018Zhang}, and we chose to have the last SPD in the array - the one indexed $i=1$ - to be 50\% absorptive as the second-last so that $L_1=L_2=\tfrac{1}{\alpha_c}\ln2$. With such hypothesis we obtain a probability of having no dark counts per measurement in the worst DCR scenario of $p(0,10\text{~ns})\simeq0.9972$.\\
Despite its flexibility, this configuration poses challenges in bending the nanowires over the waveguide. Indeed, curved nanowires often lead to current crowding issues \citep{2022Meng}, making straight sections practically ineffective. To mitigate this occurrence, widening the nanowire in curved sections is a potential solution, but it introduces absorptive \textit{dead} zones atop the waveguide, as noticeable in Fig.~\ref{fig:Cross_Section}b, significantly impacting the fidelity.

An alternative approach involves arranging the nanowires transversally to the waveguide in a meander structure, as showed in Fig.~\ref{fig:Different_configurations}b.
This arrangement places the curved parts of the nanowires farther from the waveguide, significantly reducing the absorption of the \textit{dead} sections.
On the other hand, as the SPD-waveguide interaction now occurs over a discrete number of nanowire elements (only those being transversal to the waveguide), the coupling coefficients of the SPDs can only be selected on a \textit{discretized} grid composed by multiples of the single-element efficency $\eta_{\text{se}}$, being defined as the absorption given by 2 \textit{active} sections. We neglect the absorption of the \textit{dead} sections as they can be optimized to be negligible, as we will show in the following.

On the basis of pros and cons of longitudinal and meander configurations, and given the results obtained by \citep{2022Meng} with a similar device, we want to discuss a third geometry. As showed in Fig.~\ref{fig:Different_configurations}c, one can fabricate the NbN SPD with a curvature maintaining shape, only changing periodically the rotation orientation like a \textit{wavy wall}. In such a way, one can in principle eliminate the current crowding issue, as here the whole nanowire has the same curvature and consequently eliminating \textit{dead} sections, and still being able to tune the efficiency of each NbN in a continuous way. On the other hand, the \textit{wavy wall} nanowire fabrication feasibility could be limited due to resolution constraints and longer processing times, but we strongly believe that it could be beneficial for a better tuning of SPDs efficiencies.

\begin{figure}[h]
    \centering
    \includegraphics[scale=.5]{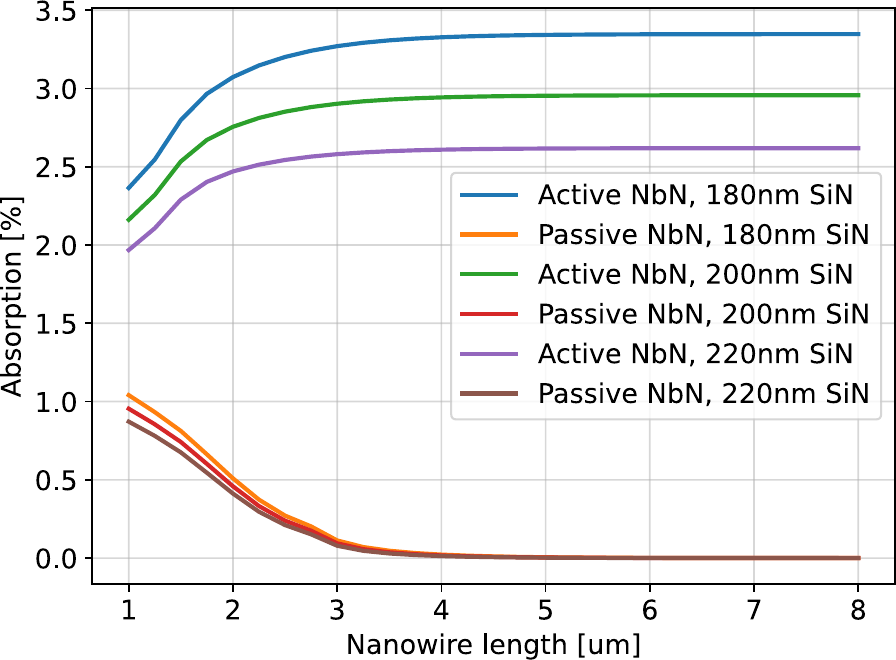}
    \caption{FEM simulation of the absorption as a function of the nanowire length. The latter takes into account both \textit{active} (transversal straight) and \textit{passive} (block for the curve) of a NbN nanowire of fixed $6$~nm thickness and $80$~nm width. The parametric sweep was performed on three waveguide heights and with nanowire transversal lenght ranging from 1~$\mu$m to 8~$\mu$m with 0.25~$\mu$m steps.}
    \label{fig:Simulation_results}
\end{figure}

Due to the better ease of the fabrication process, we opt to proceed our study with a PNRD composed of meander SPDs.
To evaluate $\eta_{\text{se}}$, we conducted FEM simulations of both \textit{active} and \textit{passive} regions, sweeping both waveguide height and NbN nanowire transversal length. The latter parameter is the defined as the joint length of \textit{active} and \textit{passive} parts of the nanowire meander, hence comprising a straight section and the curve.
The simulation results of this study are represented in Fig.~\ref{fig:Simulation_results} as absorption percentages versus the transversal length. 
Notably, both regions saturate their absorption over a length of approximately  5~$\mu$m, with the active portion reaching saturation between 2.5 and $3.5\%$, depending on the SiN thickness, while the dead zone exhibits
absorption below $6\cdot10^{-3}\%$.

As previously mentioned, the \textit{dead} section of the nanowire can be considered to have negligible absorbance, resulting in a single-element SPD efficiency for a 200~nm thick SiN waveguide of $\eta_{\text{se}}=2.957\%$.
To determine the index of the smallest feasible element in the array, we can invert Eq.~\ref{Equation:eta_i} and solve for $N$, given by:
\begin{equation}
    N=\left\lfloor1/\eta_{\text{se}}\right\rfloor    
\end{equation}
where $\left\lfloor\boldsymbol{\cdot}\right\rfloor$ denotes the floor function, representing the integer part of the expression. In our implementation, this calculation yields to $N=33$ SPDs for the PNRD.

The efficiency of the $i$-th SPD can be discretized using the following equation:
\begin{equation}
    \eta_i = 1 - \left(1-\eta_{\text{se}}\right)^{u_i}
    \label{eq:discretization}
\end{equation}
where $u_i$ denotes the number of units required to achieve the efficiency $\eta_i$ using $\eta_{\text{se}}$.
The corresponding $u_i$ values for each SPD can be determined by inverting Eq.~\ref{eq:discretization}
\begin{equation}
    u_i = \operatorname{round}\frac{\log\left(1-\frac1i\right)}{\log\left(1-\eta_{\text{se}}\right)}
\end{equation}
where the round operator is employed because $u_i$ must be an integer.

The values obtained for each $i$ between 1 and 33 are reported in Tab.~\ref{tab:Meanders}.
\renewcommand{\arraystretch}{1.5} 

\begin{table}[h]
    \centering
    \begin{tabular}{||c|c||c|c||c|c||}
        \hline
        index $i$ & $u_i$ & index $i$ & $u_i$ & index $i$ & $u_i$\\
        \hline
        33 & 1 & 22 & 2 & 11 & 3 \\
        32 & 1 & 21 & 2 & 10 & 4 \\
        31 & 1 & 20 & 2 & 9 & 4 \\
        30 & 1 & 19 & 2 & 8 & 4 \\
        29 & 1 & 18 & 2 & 7 & 5 \\
        28 & 1 & 17 & 2 & 6 & 6 \\
        27 & 1 & 16 & 2 & 5 & 7 \\
        26 & 1 & 15 & 2 & 4 & 10 \\
        25 & 1 & 14 & 2 & 3 & 14 \\
        24 & 1 & 13 & 3 & 2 & 23 \\
        23 & 1 & 12 & 3 & 1 & 77 \\
        \hline
    \end{tabular}
    \vspace{5pt}
    \caption{List of $u_i$ meanders of each SPD}
    \label{tab:Meanders}
\end{table}

\begin{figure}[t]
    \centering
    \includegraphics[scale=0.5]{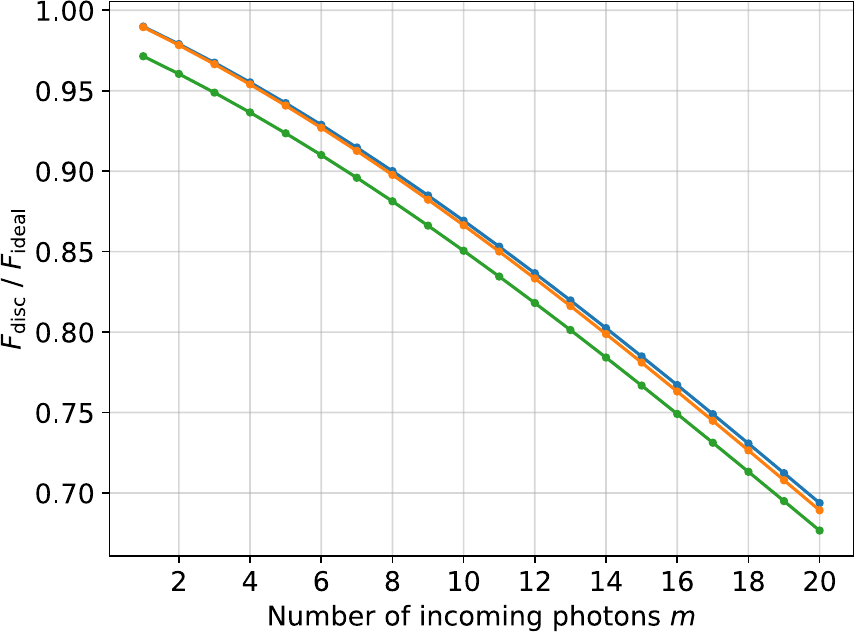}
    \caption{Performance of the discretized array obtained as the discretized fidelity normalized with the ideal fidelity. The three lines represent the different regimes: discretized only (blue line), discretized with propagation losses (orange line) and discretized with both propagation losses and dark count (green line) all considering 33 SPDs in the PNRD. The DC contribution is $p\left(0,10\text{~ns}\right)=98.17\%$.}
    \label{Figure:PerformanceComparison}
\end{figure}

Using these data and the methodologies discussed in Sec.~\ref{sec:Performance} and the worst dark counts scenario in \citep{2019Gaggero} we evaluate the performance of a PNRD composed by $N=33$ meander SPDs in different regimes: ideal, lossy and lossy plus DC. We then compare these results to the non-discretized ideal scenario by examining the fidelity ratio ${F_{\text{disc}}}/{F_{\text{ideal}}}$, as shown in Fig.~\ref{Figure:PerformanceComparison}: remarkably, the introduction of losses did not change significantly the performance of such PNRD, which the introduction of dark counts did instead.

Finally, we explicitly report the fidelities from 1 to 4 photons of such a detector considering both losses and dark counts: $F(1,33)=98.95\%$, $F(2,33)=94.87\%$, $F(3,33)=88.04\%$ and $F(4,33)=79.00\%$.

\section{Discussion and conclusions}
\label{sec:Discussion}
Performance and design considerations of a linearly multiplexed photon number-resolving single-photon detector array have been discussed. Our investigation focused on analyzing the fidelity of the PNR in detecting $m$ incident photons under various conditions and proposing a practical design for its implementation.

We began by deriving expressions for the fidelity of the SNSPD array, starting from the ideal case and then considering perturbation factors such as propagation losses and dark counts. Through theoretical analysis and numerical simulations, we demonstrated how these factors affect the fidelity of the system.

Moving on to the design aspect, we proposed a strip-loaded waveguide design for the SNSPD array, utilizing LNOI as the platform material. By conducting FEM simulations, we evaluated the performance of a 200~$\mu$m thick SiN strip loading a 300~$\mu$m LN slab layer, including straight and curved propagation losses, the latter to understand the limitations of curved designs. We also discussed how to prevent slab-mode leakage in strip-loaded waveguides and showed that it is not an eventuality in our design.

Our investigation revealed important insights into the trade-offs involved in the design of the SNSPD array. We briefly mentioned current crowding in curved nanowires and suggested two different solutions: the longitudinal and the \textit{discretized} ones. We also studied the efficiency of the superconducting NbN nanowire detectors for different lengths and nitride thicknesses. We conclude that the absorption of both \textit{active} and \textit{passive} sections of the nanowires reach their plateau value after about 5~$\mu$m of length.
Additionally, we would like to propose another nanowire configuration, that we called \textit{wavy wall}: here, every SPD is formed by a nanowire that continuously curve along the waveguide, periodically changing the curvature orientation like a sinusoid centered on the waveguide. The advantage of such a configuration is that, in principle, it would not suffer by current crowding as it is formed by a uniform curve. 

In conclusion, our study provides valuable insight into the performance and design of linearly multiplexed photon number-resolving single-photon detector arrays. By addressing key challenges and proposing practical solutions, we aim to contribute to the advancement of quantum technologies relying on photon detection and quantum state engineering.

\section*{Acknowledgements}
L.L., F.Mar., T.H.D., A.G., F.C., A.S, F.Mat. and M.B. acknowledge the support of the PNRR MUR project PE0000023-NQSTI (Italy).

This work was funded by INFN through the CSN5-UNIDET project.

\bibliographystyle{elsarticle-num-names}
\bibliography{Limongi_Bibliography}

\end{document}